\documentclass[sigconf,authorversion,nonacm]{acmart}
 \setcopyright{none}
\AtBeginDocument{%
  \providecommand\BibTeX{{%
    \normalfont B\kern-0.5em{\scshape i\kern-0.25em b}\kern-0.8em\TeX}}}

\acmYear{2021}

\makeatletter
\def\@copyrightspace{\relax}
\makeatother

\begin{document}

\title{Can Explainable AI Explain Unfairness? A Framework for Evaluating Explainable AI}




\author{Kiana Alikhademi}
\authornote{First three authors contributed equally to this research.}
\email{kalikhademi@ufl.edu}
\affiliation{%
  \institution{University of Florida}
  \streetaddress{432 Newell Dr}
  \city{Gainesville}
  \state{FL}
  \country{United States}
  \postcode{32611}
}
\author{Brianna Richardson}
\authornotemark[1]
\email{richardsonb@ufl.edu}
\affiliation{%
  \institution{University of Florida}
  \streetaddress{432 Newell Dr}
  \city{Gainesville}
  \state{FL}
  \country{United States}
  \postcode{32611}
}

\author{Emma Drobina}
\authornotemark[1]
\email{edrobina@ufl.edu}
\affiliation{%
  \institution{University of Florida}
  \streetaddress{432 Newell Dr}
  \city{Gainesville}
  \state{FL}
  \country{United States}
  \postcode{32611}
}

\author{Juan E. Gilbert}
\email{juan@ufl.edu}
\affiliation{%
  \institution{University of Florida}
  \streetaddress{432 Newell Dr}
  \city{Gainesville}
  \state{FL}
  \country{United States}
  \postcode{32611}
}






\renewcommand{\shortauthors}{anonymous, et al.}

\begin{abstract}
Many ML models are opaque to humans, producing decisions too complex for humans to easily understand. In response, explainable artificial intelligence (XAI) tools that analyze the inner workings of a model have been created. Despite these tools' strength in translating model behavior,  critiques have raised concerns about the impact of XAI tools as a tool for `fairwashing` by misleading users into trusting biased or incorrect models. In this paper, we created a framework for evaluating explainable AI tools with respect to their capabilities for detecting and addressing issues of bias and fairness as well as their capacity to communicate these results to their users clearly. We found that despite their capabilities in simplifying and explaining model behavior, many prominent XAI tools lack features that could be critical in detecting bias. Developers can use our framework to suggest modifications needed in their toolkits to reduce issues likes fairwashing.
\end{abstract}




\maketitle

\section{Introduction}

Today, machine learning and deep learning are vital to cutting-edge technologies in nearly every domain. By automating processes and removing humans from the equation, many ML practitioners aim to speed up processes, increase system performance by increasing accuracy, and reduce human bias. Recent failings of AI tools \citep{googleAIRacist, wilson2019predictive}, however, have reflected the complex nature of human bias and how it can influence ML. Due to the rise in ML users and applications, there has also been an increase in scrutiny of ML and its applications in terms of  transparency, explainability, and fairness.

Because deep ML models are difficult for humans to understand, tools to explain models and evaluate fairness have become widely discussed. While the functionality of explainable AI and fair AI tools may differ, their goals and expectations overlap considerably \citep{Abdollahi2018}. However, some explainable AI tools have received  criticism on the basis that misusing them can lead to misleading explanations that can be used to justify incorrect models \citep{Rudin}. Furthermore, in a survey study conducted by \citeauthor{Holstein}, ML practitioners depicted a strong need for tools and metrics they could use to account for and consider fairness pro-actively while building and evaluating their algorithms.  Many of the current issues of \textit{fairwashing} \citep{aivodji2019fairwashing} in Explainable AI (XAI) can be confronted using guidelines rooted in fairness ML \citep{Hall, aivodji2019fairwashing}. Moreover, another study by \citeauthor{brennen2020people} indicated that ML stakeholders want XAI that can detect bias issues. Therefore, as issues of fairness must be explained, a complete explainability tool should also deduce unfairness. To address the needs of both XAI and Fair AI tools, we propose that XAI tools include features for explaining fairness in machine learning models and data. 

We defined a rubric that can be used to create and evaluate XAI tools in terms of their functionality with respect to fairness. This work differs from fairness audits in that it evaluates and expands upon the existing functionalities of explainable AI. While other works solely focus on explainability or fairness, this is the first of its kind to fuse those functionalities into one tool. Our rubric was generated with the hope of creating a Fair Explainability toolkit that can be used across the diverse range of ML practitioners in evaluating their models and data with respect to fairness. Our major contributions are as follows:
\begin{itemize}
    \item We developed a holistic fairness rubric outlining capabilities and expectations of XAI.
    \item We used our rubric to examine several XAI tools.
    \item We outlined common gaps in functionality in regards to fairness, interpretability, and usability within these tools.
\end{itemize}

\section{Background}
\subsection{Fairness}
Mitigating the bias and unfairness within the training data is a necessity, both out of ethical duty and because of the impact that perceived inaccuracies have on user trust \citep{yin2019understanding, Toreini2020}. While the exact ways ML systems are subject to current discrimination laws are complex \citep{bornstein2018antidiscriminatory}, ML developers should consider both legal and ethical obligations to people who may be affected by their systems. However, before we can determine if a system is fair, we first need to define fairness, a surprisingly complex term. Broadly speaking, there are two paradigms of fairness: statistical (or group) and individual \citep{chouldechova2018frontiers, binns2019apparent}. 

Under the statistical definition of fairness, minority groups as a whole should be treated the same as majority groups as a whole. The parity of a selected statistical measure between these groups is the key to preserving statistical fairness. 
However, different statistical measures may give contradictory results with respect to fairness, and designers are limited in the number of measures they can choose to optimize for \citep{chouldechova2018frontiers}. Additionally, statistical fairness is limited by its need for clearly defined categories \citep{binns2019apparent}.

On the other hand, individual fairness compares each pair of individuals according to specific sets of criteria \citep{dwork2012fairness}. Essentially, "similar individuals should be treated similarly." This notion can be more intuitive for humans to understand, but detailed assumptions about measuring the similarity of instances and providing similar results must be made.

In this work, we leave our framework open for users to select and use the fairness definition and metrics of their choice. We do, however, insists that designers consider the diverse definitions of fairness and the diverse needs of the users when engaging with those tools. For the example case studies in this work, we will be considering unequal treatment of both individuals and groups and we will evaluate how well each XAI depict each.

\subsection{Explainable Artificial Intelligence}
The ability of models to provide the information necessary to verify its output is referred to as explainability. Explainability can also be important to ensuring that models comply with legal regulations and improve model accuracy \citep{samek2017explainable}. We define an explanation as a complete and accurate description of how a model generated its output. Explanations can be local (for a single instance of a decision), or global (for understanding the model's decision-making algorithm generally).

The importance of this principle has led to research efforts in industry and academia towards developing tools to help humans understand the behavior of black-box models. These tools are often referred to as explainable AI (XAI). In addition in our study, we referred to three XAI tools: LIME, AI Explainability 360, and ad-hoc explainability tools \citep{ribeiro2016should, bellamy2018ai, arya2019explanation, Breiman2001}.

\section{Rubric}

This work's main contribution is a rubric designed to assist XAI developers in building and evaluating XAI based on its ability to elucidate issues of fairness. This rubric is intended to help developers of XAI tools understand user needs; to help ML developers as they review their models for accuracy and fairness; and to help lay users critique the results of ML models \citep{wang2019designing}.  

We identified three major areas of need based on what literature suggests should be taken into account: 1) identifying issues of biased data and data processing \citep{veale2017fairerml, barocas-hardt-narayanan}; 2) reviewing the selection and optimization of the ML model under observation \citep{veale2017fairerml}; and 3) ensuring that the results of the XAI tool are useful and presented in a clear and comprehensible manner for any type of user. This section will focus on why these areas are frequently problematic and the ways XAI can assist users in identifying these issues.

\subsection{Issues with Biased data} 

One of the most well-known and supported arguments for fair ML is the acknowledgment that poor quality data will produce poor quality results. During the process of creating, collecting, and processing data, there are many avenues where human bias can be introduced into the model \citep{veale2017fairerml, barocas-hardt-narayanan, Barocas2018, Hardt2014, Corbett-Davies2018}. Selection bias can prevent data from being representative and diverse; intrinsic and measurement bias can prevent accurate data; and confirmation bias can prevent an auditor from catching inaccuracies. 

Machine learning algorithms are products of their data and data is often reflective of society, which inevitably includes society's bias. The influences of society's bias work as a feedback loop, creating data that is reflective of the discrimination and the impact of stereotypes \citep{barocas-hardt-narayanan}. For example, if personal biases prevented a bank owner from giving minorities loans, there would be insufficient evidence in the data to prove minorities could pay off loans. While some discrepancies like this might be easy to decipher, the long-term impact of systematic racism, sexism, etc. can be difficult to trace. With the extensive history of bias, inevitably, it will influence much of the available data.

Furthermore, the data collection stage is far from objective. There are biases correlated to the attributes that are decided on, the target variable that is selected, and the samples that are chosen to be included. Issues like sample size disparity exists, a seemingly intractable dilemma where insufficient data produces inaccurate models and minorities are inherently plagued by the fact that their available data is minimal\citep{Hardt2014}. Furthermore, data collectors hold the power of deciding the composition of the data set. Their decision on which attributes are important when considering the target variable is highly dependent on their own way of life and professional preferences, which may not be an appropriate determinant for those of other backgrounds \citep{veale2017fairerml, Holstein}. 

Finally, once the data is being used to create a machine learning model, it must be processed. Human biases can also influence this final stage before the creation of the algorithm. Steps like 'feature engineering', are completely subjective but highly influential to the final algorithm \citep{veale2017fairerml}. The choice of how to handle imperfect or missing data is also at the ML engineer's discretion and can have drastically different results on the algorithm \citep{Garcia2016}. 

While an XAI tool cannot know how the data was collected, the raw and processed data's availability can allow an XAI tool to reveal imbalances that may exist. Considering the limitations of the XAI, we identified four major areas where an XAI tool could assist in identifying issues of biased data:
\begin{itemize}
    \item The XAI tools could identify imbalances within the data as it relates to over/under-sampling;
    \item The XAI tools could identify attributes most influential in both local and global decisions;
    \item The XAI tools can identify processing issues that had a distinct impact on the final model;
    \item The XAI tools can consider the impact of user-labeled sensitive attributes on the model performance.
\end{itemize}

\subsection{Issues with the Selection \& Creation of ML Models}

Just as the biases of the data affect the resulting algorithm, the selection of ML models and the constraint functions used to optimize the algorithm can also introduce their own unintended influences into the results \citep{veale2017fairerml}.

To start, the selection of a model can impact the quality of the results (and the explanations that correspond). For example, the relationship between variables is lost in regression models, where each variable is considered independently of all other variables \citep{veale2017fairerml}. Furthermore, if selection bias exists within the data, some models, like regression and Bayesian classifiers, are built to better handle such data \citep{Zadrozny2004}. Even more generally, the selection of a family of methods may be incorrect. Often when given labeled data, the assumption is that the labels given are correct, so supervised methods are used. However, it may be more appropriate to utilize unsupervised methods prior to, in tandem with, or in place of supervised methods \citep{Benthall2019}.

The functions used to optimize the machine learning models during the creation can also introduce some fairness issues into the model. Just as the lack of penalty functions can lead to overfitting, failure to check for fairness metrics like classification parity and calibration can result in an algorithm that can be labeled as unfair \citep{Corbett-Davies2018}. Furthermore, evaluation metric's selection may be a reflection of the biases or motivators that the stakeholders have, such as maximizing profit or minimizing client risk.

In addition, the selection of evaluation and optimization functions, especially with respect to fairness, is wide and choosing the best one is largely situation-dependent \citep{Gajane2018}. It is important when considering fairness that the user can select their own evaluation metric.

Considering these fairness issues, we highlighted two general expectations for any fairness XAI:
\begin{itemize}
    \item The XAI tools can highlight influences from model selection and optimization that impacted the final algorithm and it's performance and
    \item The XAI tools consider some metric of fairness in evaluating the global performance of the resulting algorithm.
\end{itemize}

\subsection{Issues involved with the presentation of XAI results}

The final section of the rubric considers the wide variety of users that should be considered when making XAI and fairness technology. Here, XAI tools are evaluated based on how complete and intuitive their explanations are. If either from the lack of statistical background or the employment of 'black-box' models, ML practitioners employ XAI as a tool to get a better understanding both locally and globally of the behavior of their model. Furthermore, fairness tools must be built around the audience they are meant for. Machine learning experts, stakeholders, and consumers all use the technology with their own expectations of what to gain with respect to explainability \citep{ras2018explanation} and fairness. It is also important that a variety of explanation types are provided to supplement the variety of fashions to which unfairness can slip into the algorithm \citep{Dodge2019}.

As \citeauthor{ribeiro2016should} found when implementing LIME, it was important to optimize usability features to allow all users to understand and trust, gain insights about, and improve their model \citep{ribeiro2016should}.

Therefore, the final portion of the rubric allows the XAI developers to evaluate how successful their tool is as a usable device:
\begin{itemize}
    \item Does the XAI tool clearly identify its target audience and their expectations for the tool?
    \item Is the presentation of explanations sufficient for the target audience to gain insight and improve upon their model?
    \item Does the XAI tool provide a variety of types of explanations?
\end{itemize}
Using this set of expectations, we evaluated the four XAI tools. The study methodology can be found in the following sections.

\section{Experiments}
\subsection{Dataset}
Through our study we used the COMPAS dataset released by ProPublica in 2016 \citep{larson2016we}. This dataset is well known as an example of a biased dataset which can showcase the main contributions of our work. COMPAS data consists of 6167 records and eight features such as age, gender, race, prior history of arrest and, etc. Decile score is the label to predict the chance that an individual would commit the crime again \citep{larson2016we}. According to ProPublica's study \citep{angwin2016machine}, there are multiple instances of racial and gender bias in this dataset \citep{angwin2016machine}.


\subsection{Machine Learning Models}
We  implemented random forest, logistic regression, and deep learning models on the COMPAS dataset. Logistic regression and random forest models were developed using the scikit-learn library. For both implementations, default settings were used with no parameter optimization. The logistic Regression model yielded a performance of 74.52\%, while Random Forest's model scored 71.88\% on the test data.\\
\indent The model has been trained for 100 epochs with batch sizes of 10. We modeled deep neural network after the models built by ProPublica when they first detected bias in the COMPAS data set \citep{larson2016we, angwin2016machine}. The deep neural network model was built using the sequential model in Keras API \citep{tensorflow}. We used ADAM optimizer and binary cross-entropy as optimizer and loss function, respectively. The deep neural network model reached a 73.76\% accuracy. 

\section{Discussion}

Our evaluation of the four XAI tools can be seen in Table \ref{rubric}. It is important to note that this work does not evaluate how well each tool performs in the respective category, but rather whether or not the respective tool has a feature that satisfies that requirement. This section discusses the strengths and weaknesses of each XAI tool as they relate to our metrics of fairness. 
\begin{table*}
\caption{Final rubric with XAI tools being evaluated based on whether they did (+) or did not (-) include at least one example of these fairness considerations.\label{rubric}}
  \label{tab:rubric}
  \begin{tabular}{|l|c|c|c|c|}
\hline
 & \begin{tabular}[c]{@{}c@{}} \textbf{Random Forest} \\ \textbf{Feature Importance} \end{tabular} & 
 \textbf{LIME} & \textbf{AI Explainability 360} & \begin{tabular}[c]{@{}c@{}} \textbf{Ad-hoc}\\ \textbf{Explainability} \end{tabular} \\ \hline \hline
Model & Random Forest & \multicolumn{2}{c|}{Deep learning} & Logistic Regression \\ \hline
\multicolumn{5}{|c|}{\textbf{Issues with  Biased Data}} \\ \hline
Imbalanced data & - & + & + & - \\ \hline
Influential variable identification & + & + & + & + \\ \hline
Preprocessing issues & - & - & - & - \\ \hline
Sensitive attributes & - & - & - & - \\ \hline
\multicolumn{5}{|c|}{{ \textbf{Issues involved in Machine Learning Models}}} \\ \hline
Model-Specific influences & - & - & - & - \\ \hline
Accuracy equity & - & - & + & - \\ \hline
\multicolumn{5}{|c|}{\textbf{Issues involved with XAI results}} \\ \hline
Target audience & - & - & + & - \\ \hline
Presentation of explanations & + & + & + & + \\ \hline
Variety of explanations & - & - & + & - \\ \hline
\end{tabular}
\end{table*}
\subsection{Logistic Regression}
Logistic Regression (LR) is one of the simplest machine learning models, with intrinsic interpretability that allows for ad-hoc explainability. This line-of-best-fit approach to decision modeling creates a set of coefficients that define how each feature impacts global behavior. The coefficients for the LR model created using the COMPAS data can be seen in Figure \ref{fig:compas_lr_coef} in the appendix. If the user decides to, assuming they have the mathematical knowledge to do so, local explanations can be generated using the coefficients, the intercept, and the sample data. While this approach is rather easy and does explain the global behavior, it requires additional processing to properly elicit issues of bias from the explanations.

Insights from LR about COMPAS can be gathered from the coefficients shown in Figure \ref{fig:compas_lr_coef} in the appendix. From looking at these results, the user may deduce that being younger than 25 and being African American are the most influential to their chance of being labeled as ``High Risk''. On the other hand, being older than 45, having an undefined race, or having a misdemeanor charge can reduce their chance of being labeled as 'High Risk'. Ideally, in this case, these results will encourage most users to further analyze the influence of bias in their data set. However, the existence of proxies, issues with preprocessing steps or imbalanced data could easily be neglected in these explanations. 

While LR was able to identify influential variables, it did not indicate imbalanced data, issues with preprocessing steps, or options to highlight the influence of sensitive attributes. Furthermore, LR provides little to no feedback with respect to whether it is a sufficient tool for the task at hand, nor does it allow a user to easily incorporate their own fairness metric to evaluate their model. While LR did receive credit for its intuitive explanations for users with some statistical background, the need for additional processing is definitely required to elicit issues of fairness.
\begin{figure}[ht]
\centering
\includegraphics[width=0.49\textwidth]{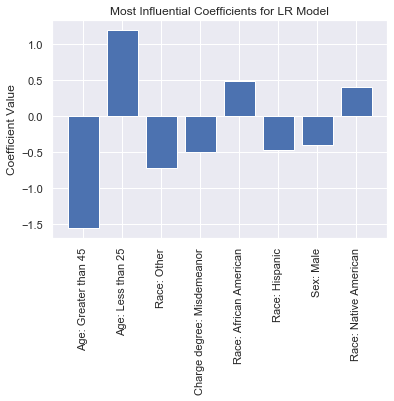}
\caption{Coefficient values for most influential attributes in Logistic Regression Model}
\label{fig:compas_lr_coef}
\end{figure}
\subsection{Random Forests}
Random Forest (RF) works as a bagged decision tree model where a forest of decision trees are constructed and decisions are determined in a 'majority rules' fashion. Random forest is popular due to its versatility, parallelization, high training speed, compatibility with high dimensionality, ability to handle unbalanced data, and low bias and variance. On the other hand, it also has multiple drawbacks, such as a low level of interpretability, high memory usage, and its reliance on parameter tuning to avoid overfitting.

Feature importance is the major feature for explainability employed by random forest. Depending on the toolkit employed, how important features are defined can be different. For our work, we employed Scikit-Learn's RandomForestClassifer \citep{scikit-learn}. Scikit-Learn defines feature importance using the methodology, known commonly as 'Gini Importance' \citeauthor{Breiman2001}. It is calculated using the impurity of a node and how often it is visited. Unlike Logistic Regression, understanding the mathematical theories behind Random Forest's explanations are a bit less intuitive. However, as can be seen in Figure \ref{fig:compas_rf_imp} in the Appendix, certain insights can be easily gathered from the feature importance. In this plot, the features are sorted from the most to the least influential. These rules are human-readable in theory, but with a large number of features and/or complex decision structures they quickly become impractical to actually interpret by scores, alone. Furthermore, the decision rules need additional evaluation to determine how they correspond to fairness.

While RF was able to identify important variables, unlike LR, it gave no indication of whether they negatively or positively impacted an individual's chance of receiving the desired result. In most other aspects, RF was similar to LR. There was little feedback with respect to biased data and model-specific issues. While the explanations were easy to access and relatively intuitive, it was difficult to access the methodology Scikit-Learn utilized to define variable importance. For Random Forest, additional processing would definitely be needed to identify issues of fairness.
\begin{figure}[ht]
\centering
\includegraphics[width=0.49\textwidth]{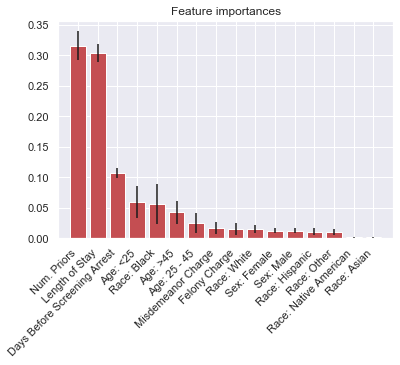}
\caption{Feature importance scores for variables in Random Forest Model}
\label{fig:compas_rf_imp}
\end{figure}
\subsection{LIME}
While the first two examples were rather simple models, explainability is most often discussed in the context of deep neural networks with numerous hidden layers and sophisticated architecture. Therefore, when analyzing LIME and IBM's AIX360, we created neural network models and used them to assess these last two XAI. 

LIME generates local explanations for black-box models by generating locally perturbed input data and investigating how model behavior changes toward this data. LIME is then able to distinguish which particular perturbations were the most influential in model predictions. Given a user-specified sample, LIME is able to provide explanations that detail the direction and degree of influence that each attribute had on that model's prediction for that sample. In Figure \ref{fig:compas_lime_exp} in the appendix, the results of employing LIME on the COMPAS neural network can be seen. Given the output shown, a user can surmise that the individual's prior offenses, length of their jail stay, and the fact that they were not over 45 led to their label as 'High Risk'. In addition to the weights of each variable, LIME defines rules that led to the attribute's weight. For this example, the number of priors being greater than 4 attributed to the higher chance of being labeled as 'High Risk'.

LIME also provides local linear approximations, detailing how changing an individual value in a sample could change the prediction probabilities. LIME's usability features exceed those of the XAI tools discussed thus far. With tabular data, it provides explanation in two formats for the user's understanding. However, it lacks a global explanation of model behavior which is critical when dealing with large complex data sets and identifying issues of fairness. It's score correlate very closely to logistic regression because they provide similar information to the user with respect to fairness. It still lacks the skills to detect issues of biased data and detect issues in the selection or processing of the model.
\begin{figure}[ht]
\centering
\includegraphics[width=0.49\textwidth]{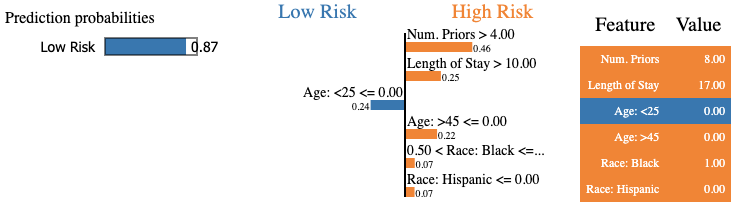}
\caption{LIME Explanation generated based on the outcome of the Neural Network model.}
\label{fig:compas_lime_exp}
\end{figure}

\subsection{IBM's AI Explainability 360}
IBM's Artificial Intelligence Explainability 360 (AIX360) is a leading toolkit in the XAI domain. Within its API, users have access to over 9 explainable methods. Within its tutorials, detailed information is provided, specifying which method may be optimal for which type of user. These explainable methods vary in type from ad-hoc to post-hoc and local to global. The goal of these methods also differ between them; some work solely on the data, simplifying complex data and highlighting nuances that were found. Other methods are used for the model, generating explanations for rules and patterns found within the specified classifiers.

AIX360 satisfied at least one requirement in every rubric category. It provides explanations and feedback on data in a variety of formats. Nonetheless, there is still room to expand upon the features AIX360 provides. For example, while AIX360 identifies outliers within the data that can hint at imbalances, it may miss larger imbalances like selection bias. 

AIX360's higher score on the rubric is no surprise considering IBM also built the Artificial Intelligence Fairness 360 (AIF360), which works towards incorporating fairness throughout the entire data processing phase. While this work focused on XAI tools, AIF360 shows promise, especially if used in tandem with AIX360.

\section {Recommendations for Future Work}

Based on our rubric categories and our experiences evaluating our models, we have several recommendations for designers of future XAI tools. 
\begin{enumerate}
    \item \textbf{Issues with Biased Data:} Allow users to select sensitive attributes that they want to focus their evaluations on. This is a key feature for explainability that has a clear purpose for evaluating fairness in regards to protected class, as well as in non-fairness-related explanations where certain features should be weighed more or less heavily in class selection than others. 
    Additionally, it is important for XAI tools to give consideration to potential issues in pre-processing data. For example, label encoding can be misused to rank categorical variables as if they were numerical variables, which XAI tools should identify by comparing the raw and processed data. As the performance of ML models is reliant on the data they receive, XAI tools should have insight into the data itself as well as the models.
    \item \textbf{Issues Involved in ML Models:} Conversely, it is important for XAI tools to evaluate the choice of model, not just the model's output. Not all models are appropriate for all data sets, and identifying such an issue can be key to uncovering the underlying problems with a model. Moreover, XAI tools should allow users to compare and contrast the distribution of predictions for subgroups against each other and against the data set as a whole. This is a vital part of understanding the global perspective on a model. 
    \item \textbf{Issues Involved with XAI Results:} XAI tools must consider their target audience and purpose. As ML permeates more areas of our society, multiple groups of people will need to evaluate ML models. These groups will have different goals and different levels of prior knowledge of ML. XAI tools with specific purposes as well as target audiences outside of ML developers will be increasingly valuable in the coming years.
\end{enumerate}

\section{Conclusion}

Our evaluations reveal that while current XAI tools provide important functions for data and model analysis, they are still lacking when it comes to analyzing fairness and could easily lead to fair-washing. This is a critical gap in XAI research, given several notable scandals in recent years in regards to bias in ML \citep{googleAIRacist, wilson2019predictive}. We hope to inspire future work into designing XAI tools that score highly on our rubric. Since ML developers and outside auditors and critics alike benefit from fairness analyses, XAI is a perfect step in the ML pipeline where we can address this need.

However, it is possible that datasets from other domains with different data types, particularly in areas such as natural language processing or time series data could be handled more or less effectively by the tools we studied. Additionally, we only investigated a limited number of XAI tools. This is one of the main areas we would like to expand upon in future work, by incorporating additional XAI tools, such as Google's What-If.

\typeout{}
\bibliographystyle{ACM-Reference-Format}
\bibliography{sample-sigconf}
\end{document}